\documentclass[epjCONF]{svjour}
\usepackage{graphics,graphicx,amsmath,amssymb,cite}
\usepackage[varg]{txfonts} 
\usepackage[latin2]{inputenc}
\usepackage[magyar,english]{babel}

\newcommand{\td}[2]{\frac{\mathrm{d}{#1}}{\mathrm{d}{#2}}}
\newcommand{\z}[1]{\left({#1}\right)}
\newcommand{\sz}[1]{\left[{#1}\right]}
\newcommand{\kz}[1]{\left\{{#1}\right\}}
\newcommand{\rec}[1]{\frac{1}{#1}}
\newcommand{\m}[1]{\mathrm{#1}}

\renewcommand{\v}[1]{\mathbf{#1}}
\renewcommand{\r}[1]{(\ref{#1})}
\newcommand{\Eq}[1]{Eq.\@ (\ref{#1})}

\session-title{International Conference on New Frontiers in Physics}

\begin{document}
\title{Time evolution of the sQGP with hydrodynamic models}
\titlerunning{Time evolution of the sQGP with hydrodynamic models \dots}
\author{M.~Csanád\inst{1}\fnmsep\thanks{\email{csanad@elte.hu}}}
\institute{Eötvös Loránd University, 1113 Budapest, Pázmány P. s. 1/a}

\abstract{
The strongly interacting  Quark-Gluon-Plasma (sQGP) created in relativistic nucleus-nucleus collisions,
can be described by hydrodynamic models. Low energy hadrons are
created after the so called freeze-out of this medium, thus their distributions reveal information
about the final state of the sQGP. Photons are created throughout the evolution, so their
distributions carry information on the whole expansion and cool down. We show results
from an analytic, 1+3 dimensional perfect relativistic hydrodynamic solution, and compare
hadron and photon observables to RHIC data.  We extract an average equation of state
of the expanding quark matter from this comparison. In the second part of this paper,
we generalize the before mentioned analytic solution of relativistic perfect fluid
hydrodynamics to arbitrary temperature-dependent Equation of
State. We investigate special cases of this class of solutions, in particular, we
present hydrodynamical solutions with an Equation of State determined from
lattice QCD calculations.
}

\maketitle

\section{Introduction}

The almost perfect fluidity of the experimentally created strongly interacting Quark-Gluon-Plasma at
the Relativistic Heavy Ion Collider (RHIC)~\cite{Adcox:2004mh} showed that relativistic hydrodynamic
models can be applied in describing the space-time picture of heavy-ion collisions
and infer the relation between experimental observables and the initial conditions.

In this paper we investigate the relativistic, ellipsoidally symmetric model
of Ref.~\cite{Csorgo:2003ry}. Hadronic observables were calculated in
Ref.~\cite{Csanad:2009wc}, while photonic observables in Ref.~\cite{Csanad:2011jq}.
We also show new solutions, which can be regarded as generalizations of the model
of Ref.~\cite{Csorgo:2003ry} to arbitrary, temperature dependent speed of sound,
originally published in Ref.~\cite{Csanad:2012hr}.

\section{Equations of hydrodynamics}

We denote space-time coordinates by $x^\mu = \z{t, \v{r}}$, with $\v{r}=(r_x, r_y, r_z)$ being the spatial
three-vector and $t$ the time in lab-frame. The metric tensor is $g_{\mu\nu}=diag\z{1,-1,-1,-1}$. Coordinate proper-time
is defined as $\tau=\sqrt{t^2-|\v{r}|^2}$. The fluid four-velocity is $u^\mu=\gamma\z{1,\v{v}}$, with $\v{v}$
being the three-velocity, and $\gamma=1/\sqrt{1-|\v{v}|^2}$. An analytic hydrodynamical solution is a functional
form for pressure $p$, energy density $\varepsilon$, 
entropy density $\sigma$, temperature $T$, and (if the fluid consists of individual conserved particles,
or if there is some conserved charge or number) the conserved number density is $n$.
Then basic hydrodynamical equations are the continuity and energy-momentum-conservation equations:
\begin{align}
\partial_\mu\z{n u^\mu} = 0\;\textnormal{ and }\;\partial_\nu T^{\mu \nu} = 0\label{e:em}.
\end{align}
The energy-momentum tensor of a perfect fluid is
\begin{align}
T^{\mu\nu} =\z{\varepsilon+p}u^\mu u^\nu-pg^{\mu \nu} .
\end{align}
The energy-momentum conservation equation can be then transformed to
(by projecting it orthogonal and parallel to $u^\mu$, respectively):
\begin{align}
\z{\varepsilon+p}u^{\nu}\partial_{\nu}u^{\mu} & =\z{g^{\mu\nu}-u^{\mu}u^{\nu}}\partial_{\nu}p,\label{e:euler} \\
\z{\varepsilon+p}\partial_{\nu}u^{\nu}+u^{\nu}\partial_{\nu}\varepsilon & = 0\label{e:energy}.
\end{align}
\Eq{e:euler} is the relativistic Euler equation, while \Eq{e:energy} is the relativistic form of the energy conservation equation. Note also that \Eq{e:energy} is equivalent to the entropy conservation equation:
\begin{align}\label{e:scont}
\partial_\mu\z{\sigma u^\mu}=0 .
\end{align}
The Equation of State (EoS) closes the set of equations. We investigate the following EoS:
\begin{align}\label{e:eos}
\varepsilon = \kappa\z{T} p ,
\end{align}
while the speed of sound $c_s$ is calculated as $c_s = \sqrt{\partial p/\partial \varepsilon}$,
i.e. for constant $\kappa$, the relation $c_s = 1/\sqrt{\kappa}$ holds.
 For the case when there is a conserved $n$ number density, we also use the well-known relation for ideal gases:
\begin{align}\label{e:tdef}
p=nT. 
\end{align}

For $\kappa\z{T}=$ constant, an ellipsoidally symmetric solution of the hydrodynamical equations is presented in Ref.~\cite{Csorgo:2003ry}:
\begin{align}\label{e:tsol0}
u^\mu = \frac{x^\mu}{\tau},\quad n = n_0\frac{V_0}{V}\nu\z{s},\quad
T = T_0\z{\frac{V_0}{V}}^{\rec{\kappa}}\rec{\nu\z{s}} ,\quad V = \tau^3,\quad
s = \frac{r_x^2}{X^2} + \frac{r_y^2}{Y^2} + \frac{r_z^2}{Z^2},
\end{align}
where $n_0$ and $T_0$ correspond to the proper time when the arbitrarily chosen volume $V_0$
was reached (i.e. $\tau_0 = V_0^{1/3}$), and $\nu\z{s}$ is an arbitrary function of $s$.
The quantity $s$ has ellipsoidal level surfaces, and obeys $u^\nu\partial_\nu s=0$. We call $s$ a
\emph{scaling variable}, and $V$ the effective volume of a characteristic ellipsoid.
Furthermore, $X$, $Y$, and $Z$ are the time (lab-frame time $t$) dependent principal axes of an expanding ellipsoid. They have the explicit time dependence as $X = \dot X_0 t$, $Y = \dot Y_0 t$, and $Z = \dot Z_0 t$,
with $\dot X_0$, $\dot Y_0$, $\dot Z_0$ constants.

\section{Photon and hadron observables for constant EoS}

From the above hydrodynamic solution with a constant EoS, source functions can be written up. For bosonic hadrons, it takes the following form~\cite{Csanad:2009wc}:
\begin{align}
S(x,p)d^4x=\mathcal{N}\frac{p_{\mu}\,d^3\Sigma^{\mu}(x)H(\tau)d\tau}{n(x)\exp\left(p_{\mu}u^{\mu}(x)/T(x)\right)-1},
\end{align}
where $\mathcal{N}=g/(2\pi)^3$ (with $g$ being the degeneracy factor), $H(\tau)$ is the proper-time probability distribution
of the freeze-out. It is assumed to be a $\delta$ function or a narrow Gaussian centered at the freeze-out proper-time $\tau_0$. Furthermore,
$\mu(x)/T(x)=\ln n(x)$ is the fugacity factor and $d^3 \Sigma_\mu(x)p^\mu$ is the Cooper-Frye
factor (describing the flux of the particles), and $d^3 \Sigma_\mu(x)$ is the vector-measure of
the freeze-out hyper-surface, pseudo-orthogonal to $u^\mu$. Here the source distribution is
normalized such as $\int S(x,p) d^4 x d^3{\bf p}/E = N$,
i.e. one gets the total number of particles $N$ (using $c$=1, $\hbar$=1 units). Note that one has to change variables
from $\tau$ to $t$, and so a Jacobian of $d\tau/dt=t/\tau$ has to be taken into account.

For the source function of photon creation we have~\cite{Csanad:2011jq}:
\begin{align}\label{e:source}
S(x,p)d^4x = \mathcal{N'}\frac{p_{\mu}\,d^3\Sigma^{\mu}(x)dt}{\exp\left(p_{\mu}u^{\mu}(x)/T(x)\right)-1}
= \mathcal{N'}\frac{p_{\mu}u^{\mu}}{\exp\left(p_{\mu}u^{\mu}(x)/T(x)\right)-1}\,d^4x
 \end{align}
where $p_{\mu}d^3\Sigma^{\mu}$ is again the Cooper-Frye factor of the emission hyper-surfaces. Similarly to the previous case, we assume that the hyper-surfaces
are pseudo-orthogonal to $u^\mu$, thus $d^3\Sigma^{\mu}(x) = u^{\mu}d^3x$. This
yields then $p_{\mu}u^{\mu}$ which is the energy of the photon
in the co-moving system. The photon creation is the assumed to happen from an initial time $t_i$ until a point sufficiently near
the freeze-out. From these source functions, observables can be calculated, as detailed in
Refs.~\cite{Csanad:2009wc,Csanad:2011jq}

\section{Comparison to measured hadron and photon distributions}
Observables calculated from the above source functions were compared to data in
Refs.~\cite{Csanad:2009wc,Csanad:2011jq}. Hadron fits determined the freeze-out parameters
of the model~\cite{Csanad:2009wc}: expansion rates, freeze-out proper-time and freeze-out temperature
(in the center of the fireball).
When describing direct photon data~\cite{Csanad:2011jq}, the free parameters (besides the ones fixed
from hadronic fits) were $\kappa$ (the equation of state parameter) and $t_i$, the initial time of the evolution.

\begin{table}
\begin{center}
    \begin{tabular}{ lllll }
    \hline\hline
    Dataset                	&                               	& $N_1$ and HBT	& elliptic flow    	& $N_1$\\
                                  	&                              	& hadrons            	& hadrons        	& photons\\\hline
    Central FO temperature	& $T_0$ [MeV]         	& 199$\pm$3      	& 204$\pm$7   	& $204$ MeV (fixed) \\
    Eccentricity                  	& $\epsilon$             	& 0.80$\pm$0.02 	& 0.34$\pm$0.03	& $0.34$ (fixed) \\
    Transverse expansion  	& $u_t^2/b$            	& -0.84$\pm$0.08	& -0.34$\pm$0.01	& $-0.34$ (fixed) \\
    FO proper-time            	& $\tau_0$ [fm/$c$]	& 7.7$\pm$0.1    	& -                   	& $7.7$  (fixed) \\
    Longitudinal expansion 	& $\dot{Z_0^2}/b$   	& -1.6$\pm$0.3  	& -                   	& $-1.6$ (fixed) \\
    Equation of State       	& $\kappa$               	&  -                      	& -                   	& $7.9 \pm 0.7$  \\
    \hline\hline
    \end{tabular}   
\end{center}
\caption{Parameters of the model determined by hadron and photon observables data. See details in Refs.~\cite{Csanad:2009wc,Csanad:2011jq}.}\label{t:param}
\end{table}

\begin{figure}
	\centering
		\includegraphics[height=0.32\textwidth,angle=270]{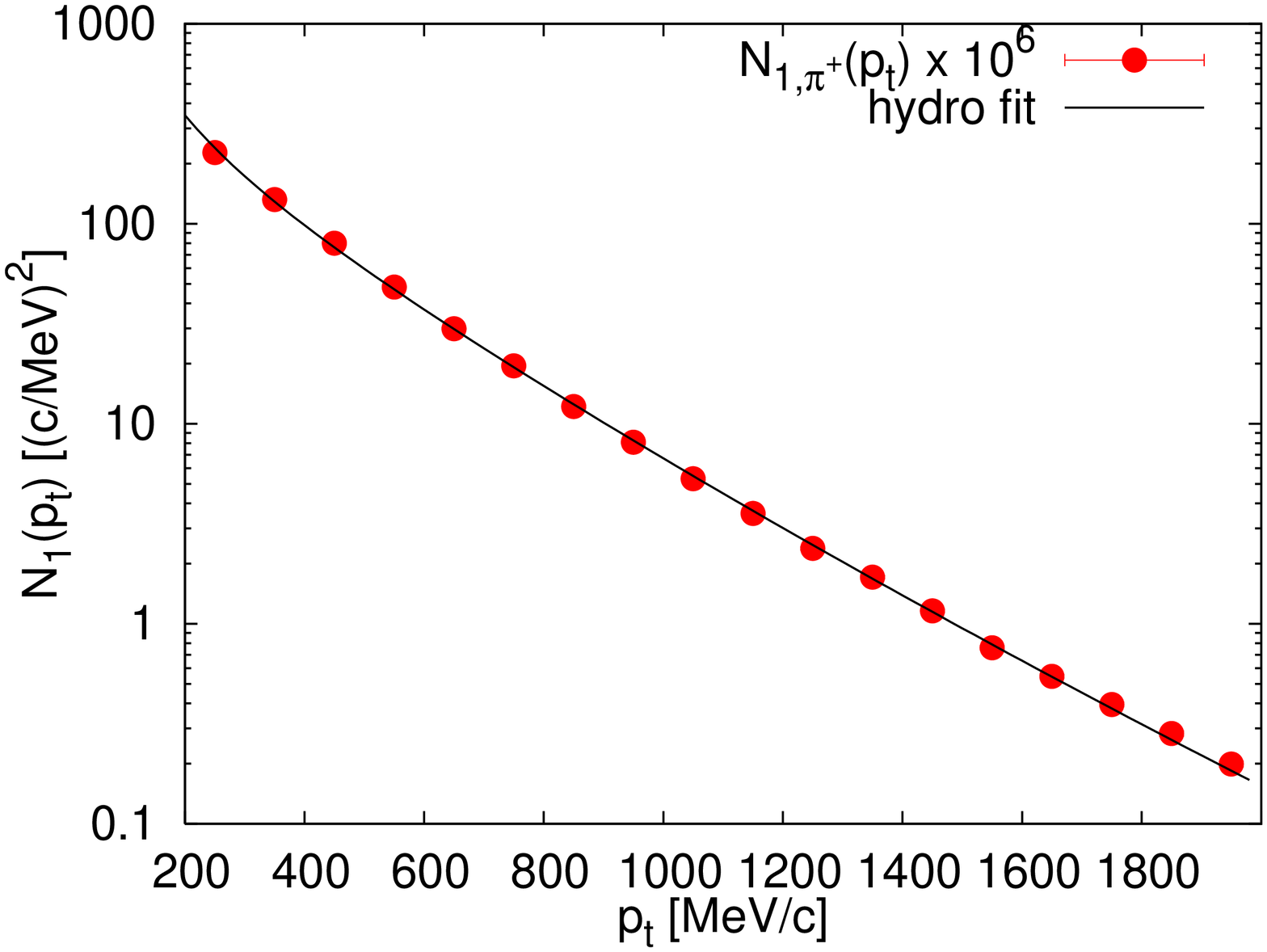}
		\includegraphics[height=0.32\textwidth,angle=270]{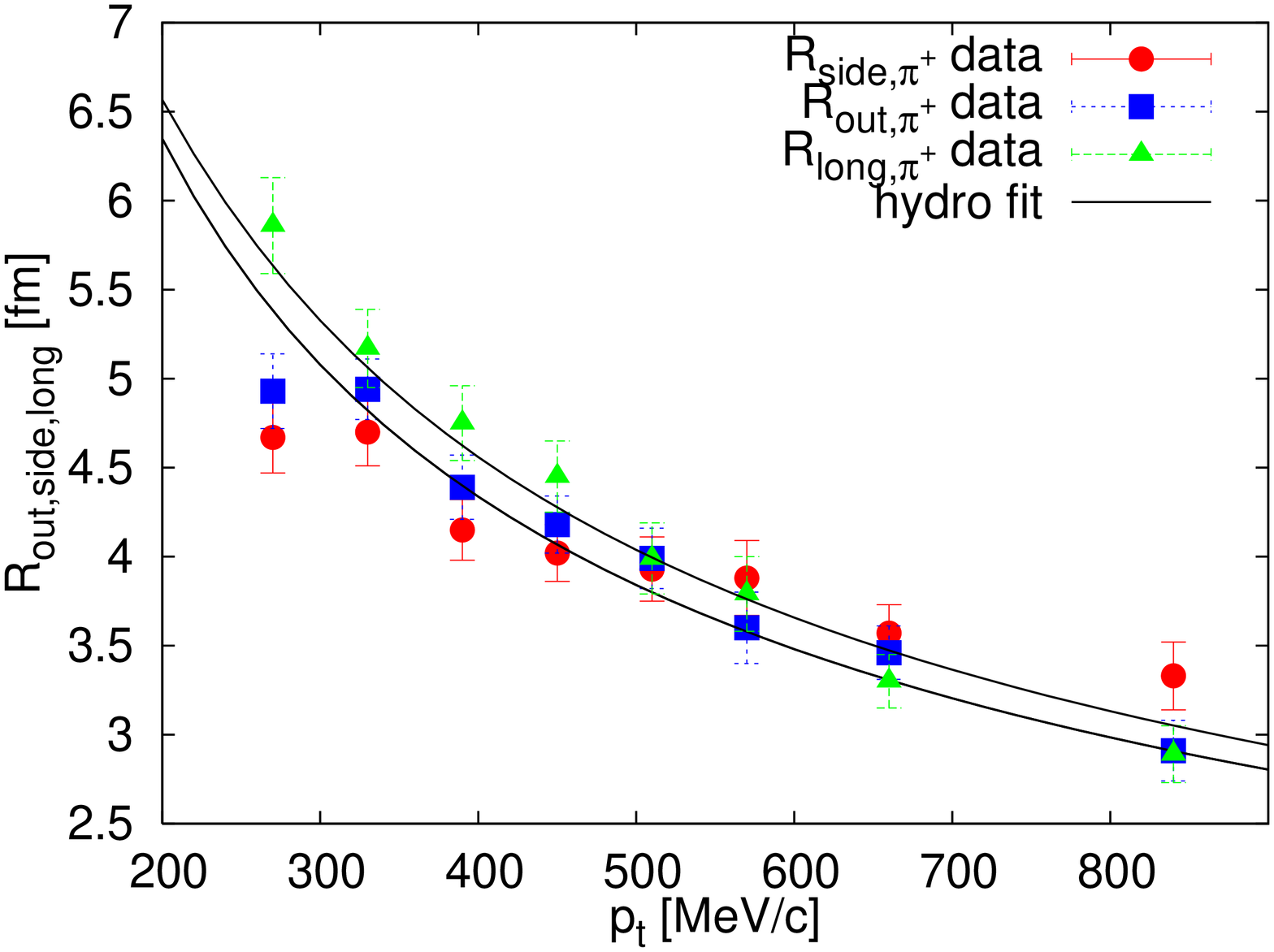}
		\includegraphics[height=0.32\textwidth,angle=270]{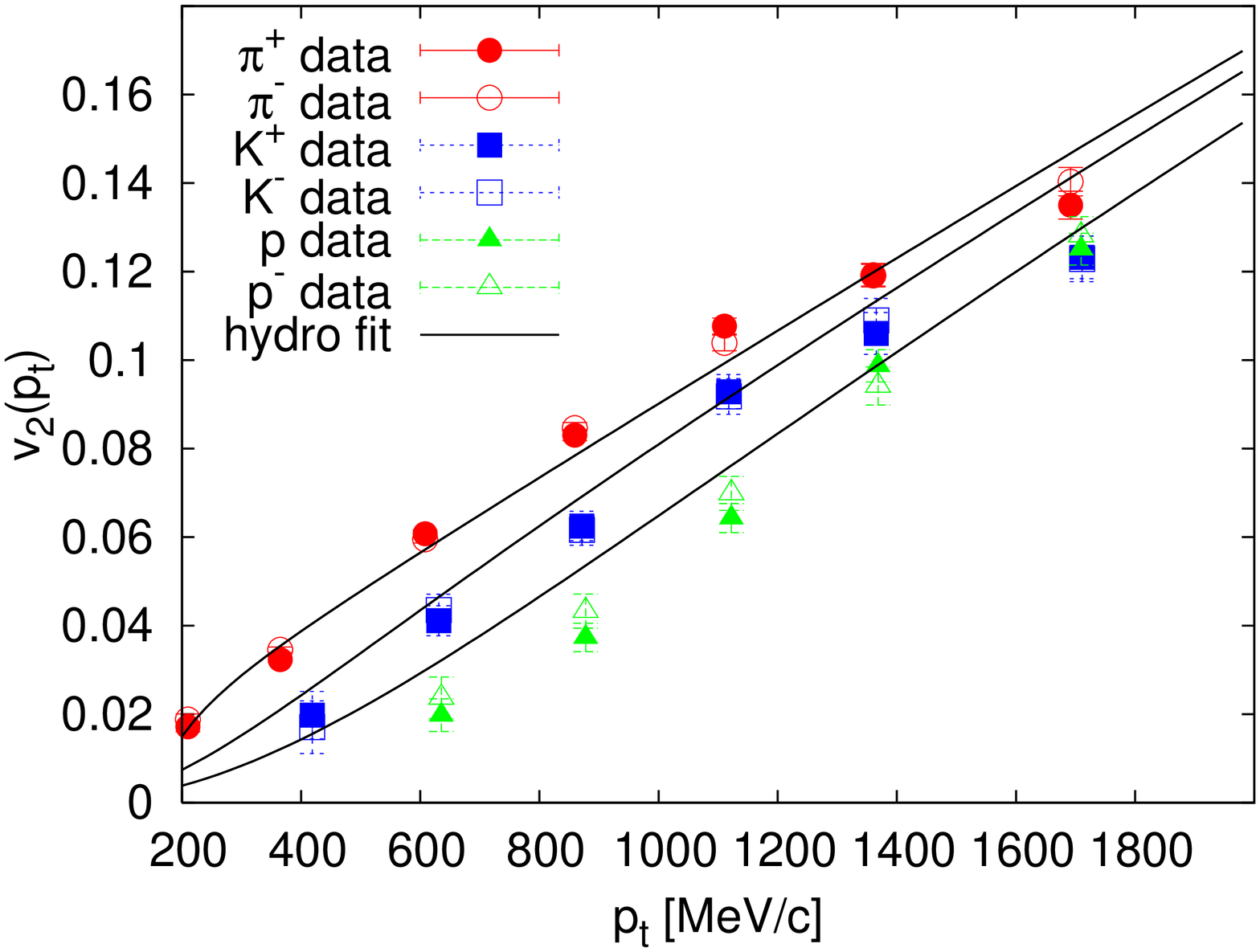}
	\caption{Fits to invariant momentum distribution of pions~\cite{Adler:2003cb} (left),
                       HBT radii~\cite{Adler:2003kt} (middle)
                         and elliptic flow~\cite{Adler:2004rq} (right). See the obtained parameters in
                          Table~\protect\ref{t:param}.}
	\label{f:hadronfits}
\end{figure}

\begin{figure}
 \begin{center}
 \includegraphics[width=0.32\textwidth]{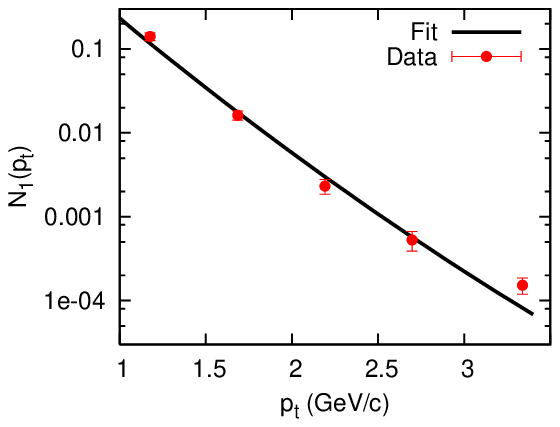}
 \includegraphics[width=0.32\textwidth]{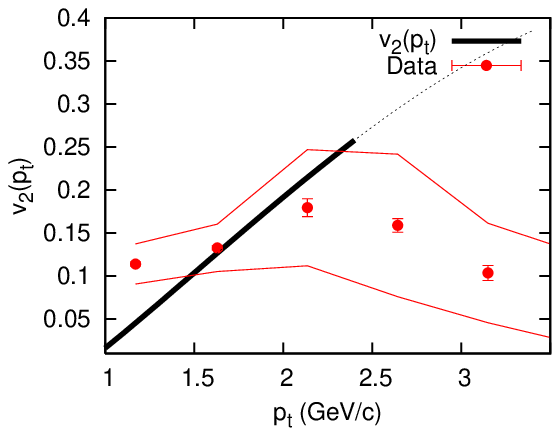}
 \includegraphics[width=0.32\textwidth]{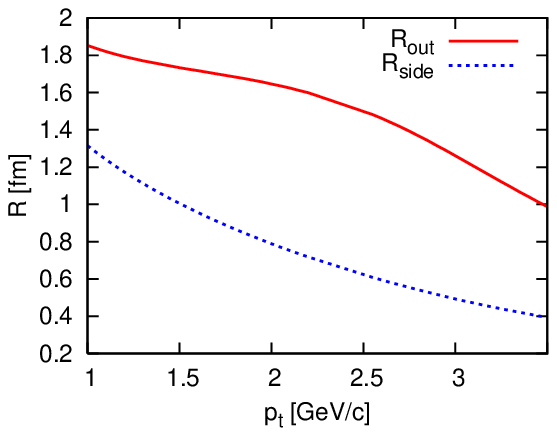}
 \end{center}
 \caption{Fit to direct photon invariant transverse momentum data~\cite{Adare:2008fqa} (left), comparison
                to elliptic flow data~\cite{Adare:2011zr} (middle) and direct photon HBT predictions (right). See the
                model parameters in Table~\protect\ref{t:param}.}\label{f:photonfits}
\end{figure}

We compared our model to PHENIX 200 GeV Au+Au hadron and photon data from Refs.~\cite{Adler:2003cb,Adler:2004rq,Adler:2003kt,Adare:2008fqa}.
Results are shown in Figs.~\ref{f:hadronfits} and~\ref{f:photonfits}, while the model
parameters are detailed in Table.~\ref{t:param}.
The EoS result from the photon fit is $\kappa=7.9\pm0.7_{stat}\pm1.5_{syst}$, or alternatively,
using $\kappa=1/c_s^2$
\begin{align}
c_s =  0.36\pm0.02_{stat}\pm0.04_{syst}
\end{align}
which is in agreement with lattice QCD calculations~\cite{Borsanyi:2010cj} and measured hadronic data~\cite{Adare:2006ti,Lacey:2006pn}. This represents an average EoS as it may
vary with temperature. The maximum value for $t_i$ within 95\% probability is 0.7 fm/$c$. The initial
temperature of the fireball (in its center) is then:
\begin{align}
T_i = 507\pm12_{stat}\pm90_{syst}\textnormal{MeV}
\end{align}
at 0.7 fm/$c$. This is in accordance with other hydro models as those values are in the $300-600$ MeV interval~\cite{Adare:2008fqa}. Note that the systematic uncertainty comes from the analysis of a
possible prefactor, as detailed in Ref.~\cite{Csanad:2011jq}.

Using the previously determined fit parameters. we can calculated the elliptic flow of
direct photons in Au+Au collisions at RHIC. This was compared to PHENIX data~\cite{Adare:2011zr},
as shown in Fig.~\ref{f:photonfits}, and they were found not to be incompatible.
We also calculated direct photon HBT radii as a prediction, and found
 $R_\textnormal{out}$ to be significantly larger than $R_\textnormal{side}$.

\section{New solutions for general Equation of State}\label{s:sols}

We found new solutions to the relativistic hydrodynamical equations for arbitary
$\varepsilon=\kappa\z{T}p$ Equation of State, as detailed in Ref.~\cite{Csanad:2012hr}.
These are the first solutions of their kind (i.e. with a non-constant EoS). In the case where we do 
not consider any conserved $n$ density, the solution is given as:
\begin{align}
\sigma &= \sigma_0 \frac{\tau_0^3}{\tau^3} ,\label{e:Tsol:s:0}\\
u^\mu & = \frac{x^\mu}{\tau} ,\\
\frac{\tau_0^3}{\tau^3} & = 
\exp\kz{\int_{T_0}^T\z{\frac{\kappa\z{\beta}}{\beta}+\rec{\kappa\z{\beta}+1}\td{\kappa\z{\beta}}{\beta}}\m{d}\beta} .
\label{e:Tsol:s}
\end{align}
For the case when the pressure is expressed as $p=nT$ with a conserved density $n$, another new solution
can be written up as:
\begin{align}
n &= n_0 \frac{\tau_0^3}{\tau^3} ,\\
u^\mu & = \frac{x^\mu}{\tau} ,\\
\frac{\tau_0^3}{\tau^3} & =
\exp\kz{\int_{T_0}^T\z{\rec{\beta}\td{}{\beta}\sz{\kappa\z{\beta}\beta}}\m{d}\beta} . \label{e:Tsol:nT}
\end{align}
Quantities denoted by the subscript 0 ($n_0$, $T_0$, $\sigma_0$) correspond to the
proper-time $\tau_0$, which can be chosen arbitrarily. If for example $\tau_0$ is taken
to be the freeze-out proper-time, then $T_0$ is the freeze-out temperature. 
These solutions are simple generalizations of the $\nu\z{s}=1$ case of the solutions of Ref.~\cite{Csorgo:2003ry},
and the latter also represents a relativistic generalization of the solution presented in Ref.~\cite{Csorgo:2001xm}.

It is important to note that the conserved $n$ solution becomes ill-defined,
if $\td{}{T} \z{\kappa\z{T}T}>0$ is not true. In such a case, one can
use the solution without conserved $n$ (Eqs.~\r{e:Tsol:s:0}--\r{e:Tsol:s}).

If $\kappa$ is given as a function of the pressure $p$ and not that of the temperature $T$,
a third new solution can be given as:
\begin{align}
\sigma &= \sigma_0 \frac{\tau_0^3}{\tau^3} ,\label{e:psol:s}\\
u^\mu & = \frac{x^\mu}{\tau} ,\\
\frac{\tau_0^3}{\tau^3} & = 
\exp\kz{\int_{p_0}^p\z{\frac{\kappa\z{\beta}}{\beta}+\td{\kappa\z{\beta}}{\beta}}\frac{\m{d}\beta}{\kappa\z{\beta}+1}} ,
\label{e:psol:p}
\end{align}
i.e. almost the same as in \Eq{e:Tsol:s}, except that here the integration variable is the pressure $p$.

\section{Utilizing a lattice QCD EoS}
\begin{figure}
 \begin{center}
 \includegraphics[angle=270,width=0.490\textwidth]{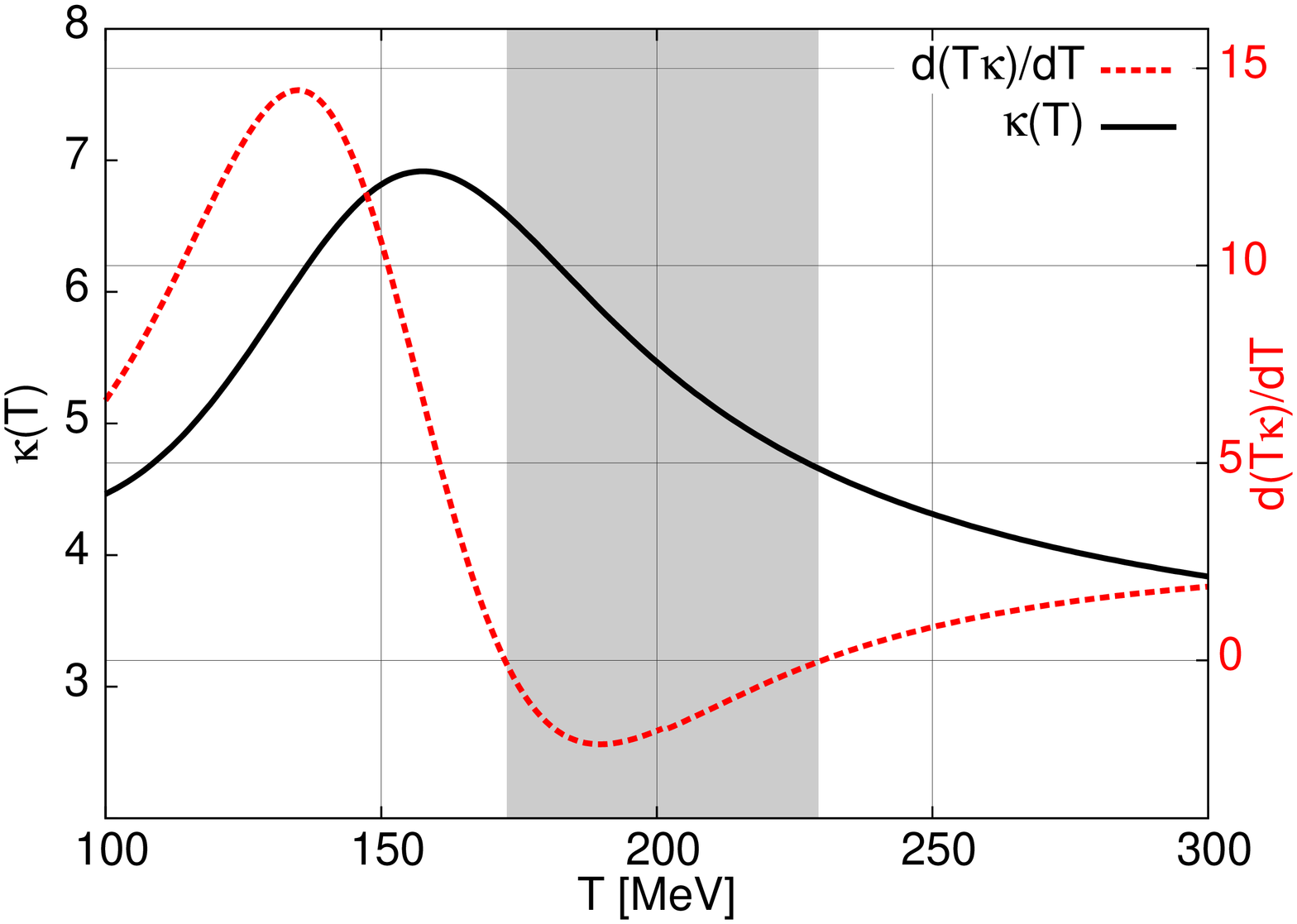}
 \includegraphics[angle=270,width=0.483\textwidth]{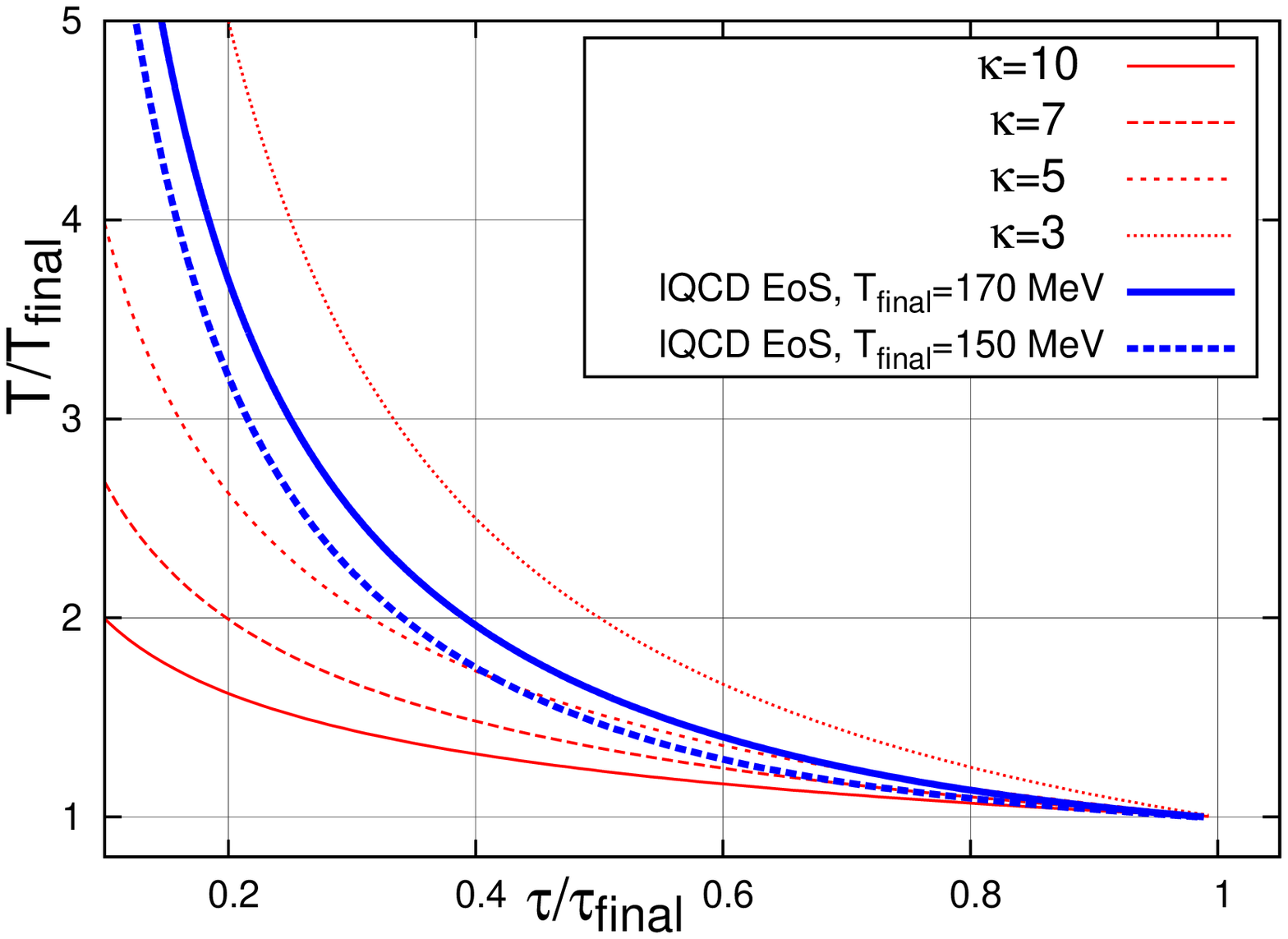}
 \end{center}
 \caption{Left: The temperature dependence of the EoS parameter
$\kappa$ from  Ref.~\cite{Borsanyi:2010cj} is shown with the solid
black curve. In the shaded $T$ range (173 MeV - 230 MeV)
$\td{}{T} \z{\kappa\z{T}T}$ (red dashed line) becomes
negative, thus the solution shown in Eqs.~\r{e:Tsol:s:0}--\r{e:Tsol:s} shall be used with this EoS.
Right: Time dependence of the temperature $T(\tau)$ (normalized with the freeze-out time $\tau_f$
and the freeze-out temperature $T_f$) is shown. The four thin red lines show this dependence in
case of constant $\kappa$ values, while the thicker blue lines show results based on the EoS of Ref.~\cite{Borsanyi:2010cj}.
}\label{f:validity_ttau}
\end{figure}

A QCD equation of state has been calculated by the Budapest-Wuppertal group in Ref.~\cite{Borsanyi:2010cj},
with dynamical quarks, in the continuum limit.
In their Eq.~(3.1) and Table 2, they give an analytic parametrization of the trace
anomaly $I=\epsilon-3p$ as a function of temperature. 
The pressure can also be calculated from it, as (if using the normalized values and $\hbar=c=1$ units)
$\frac{I}{T^4}=\frac{1}{T}\frac{\partial}{\partial T}\frac{p}{T^4}$.
From this, we calculated the EoS parameter $\kappa=I/p+3$ as a function of the temperature,
as shown in Fig.~\ref{f:validity_ttau} (left plot).
Since in a $T$ range $\td{}{T} \z{\kappa\z{T}T}$ becomes negative, the solution without
conserved number density $n$ (presented in Eqs.~\r{e:Tsol:s:0}--\r{e:Tsol:s}) was used.~\cite{Csanad:2012hr}

We utilized the obtained $\kappa(T)$ and calculated the time evolution of the temperature of the fireball from this
solution of relativistic hydrodynamics. The result is shown in Fig.~\ref{f:validity_ttau} (right plot).
Clearly, temperature falls off almost as
fast as in case of a constant $\kappa=3$, an ideal relativistic gas. Hence a given freeze-out temperature yields a
significantly higher initial temperature than a higher $\kappa$ (i.e. a lower speed of sound $c_s$) would.

Let us give an example! We shall fix the freeze-out temperature, based on lattice QCD results,
to a reasonable value of  $T_f = 170$ MeV. Let all the quantities with subscript 0 correspond the the freeze-out,
we shall thus index them with $f$.
In this case, already at  $0.3\times\tau_f$ (30\% of the freeze-out time), temperatures 2.5-3$\times$
higher than at the freeze-out can be reached. To give a full quantitative example, let
as assume the following values:
\begin{align}
\tau_f = 8\;\rm{fm}/c \;\textnormal{ and }\;&\tau_{\rm init} = 1.5\;\rm{fm}/c\textnormal{, \;then}\\
T_f = 170\;\rm{MeV}\;\Rightarrow\;&T_{\rm init} \approx 550\;\rm{MeV}
\end{align}
(and even higher if $\tau_{\rm init}$ is smaller). This value would have been reached with a constant EoS
of $\kappa\approx4$, even though the extracted average EoS values are usually above this value.
The reason for it may be, that for the largest temperature range, $\kappa$ obtains values close to 4,
as shown in the left plot of Fig.~\ref{f:validity_ttau}.

In general, the mentioned lattice QCD equation of state of Ref.~\cite{Borsanyi:2010cj} and our hydro solution
yields a $T(\tau)$ dependence. Then, if the freeze-out temperature $T_f$ and the time evolution duration
$\tau_f / \tau_{\rm init}$ are known, the initial temperature of the fireball can be easily calculated, or
even read off the right plot of  Fig.~\ref{f:validity_ttau}, as it was drawn with units normalized by
the freeze-out temperature and proper-time.

\section{Conclusion}
Exact parametric solutions of perfect hydrodynamics can be utilized in order to
describe the matter produced in heavy ion collisions at RHIC.
We calculated observables from a relativistic, 1+3 dimensional, ellipsoidally symmetric,
exact solution, and compared these to 200 GeV Au+Au PHENIX data. Hadronic
data are compatible with our model, and freeze-out parameters were extracted from fits
to these data.~\cite{Csanad:2009wc}

From fits to direct photon data, we find that thermal radiation is consistent
these measurements, with an average speed of sound of $c_s = 0.36\pm0.02_{stat}\pm0.04_{syst}$.
We can also set a lower bound on the initial temperature of the sQGP to $507\pm12_{stat}\pm90_{syst}$ MeV
at $0.7$ fm/$c$. We also find that the thermal photon elliptic flow from this mode is not incompatible with
measurements. We also predicted photon HBT radii from this model.~\cite{Csanad:2011jq}

In the second part of this paper, we have presented the first analytic solutions of the equations of
relativistic perfect fluid hydrodynamics for general temperature dependent speed of sound (ie.\ general Equation of State).
Using our solutions and utilizing a lattice QCD Equation of State, we explored the
initial state of heavy-ion reactions based on the reconstructed final state.
In $\sqrt{s_{NN}}=200$ GeV Au+Au collisions, our investigations reveal a very
high initial temperature consistent with calculations based on the measured spectrum of low momentum
direct photons.~\cite{Csanad:2012hr}

\section*{Acknowledgments}
This work was supported by the  NK-101438 OTKA grant and the Bolyai Scholarship (Hungarian Academy of Sciences)
of M. Csan\'ad.  The author also would like to thank the organizers for the possibility of participating at the first
International Conference on New Frontiers in Physics.

\bibliographystyle{epj}
\bibliography{../../../master}

\end{document}